\documentclass[11pt,a4paper]{article}
\pdfoutput=1
\usepackage[]{graphicx}
\usepackage[]{caption}
\usepackage[]{subcaption}
\graphicspath{{images/}}
\DeclareGraphicsExtensions{.png}

\title{Activity Modulation of Motor and Somatosensory Neurons in Learning}
\author{Alexander Conway \\ 
Department of Physics \\ 
The College of The University of Chicago \\ 
\and \\ 
Fritzie Arce, Ph.D \\ 
Department of Organismal Biology and Anatomy \\ 
The University of Chicago \\
\\
This Undergraduate Research Project Supported By \\ Nicholas Hatsopoulos, Ph.D \\ Department of Organismal Biology and Anatomy \\ The University of Chicago}

\begin{document}
\maketitle
\begin{abstract}
The cortical processes involved in learning are not well understood. Recent experiments have studied population-level response in the orofacial somatosensory (S1) and motor (S1) cortices of rhesus macaque monkeys during adaptation to a simple tongue protrusion task within and across multiple learning sessions. Initial findings have suggested the formation of cell assemblies during adaptation. In this report we explore differences in cell activity between successful and failed trials as the monkey learns during two sessions. The ability to directly compare data across multiple sessions is fairly new and until now research has mostly focused on the activity of neurons during successful trials only. We confirm findings of the development of coherently active cell assemblies and find that neural response differentiates significantly between successful and unsuccessful trials, particularly as the monkey adapts to the task. Our findings motivate further research into the differences in activity between successful and unsuccessful trials in these experiments.
\end{abstract}

\tableofcontents

\section{Introduction} 

\subsection{Background and Motivation}

The cortical processes involved in learning are not well understood. Recent experiments performed by the Hatsopoulos lab investigated changes in the activity of neurons in the orofacial somatosensory (S1) and motor (S1) cortices of rhesus macaque monkeys as they learned to perform a single-direction, tongue controlled targeting task over the course of several learning sessions \cite{farce-adaptation}. The S1 and M1 cortices are implicated in the control of  tongue and jaw movements. Analysis of data from these experiments has already found a number of interesting results that demonstrate the importance of a multi-session learning approach. Arce, Hatsopoulos et al. describe findings such as a decrease in trial-by-trial response variability of neurons in both cortices with a general decrease in activity modulation in the M1 cortex with skill acquisition and modulation in the coherence between spike trains and local field  potential within and across the two cortical areas across learning sessions. This latter result suggests that assemblies of coherently active cells form as a function of skill acquisition. 

In this report we use similar data to explore the differences in neural activity between successful and unsuccessful trials and how it changes across learning sessions. The ability to directly compare neural data across trials is fairly novel in this type of experiment and this is the first time that the differences in activity due to the outcome of a trial has been explored. We find marked changes in neural activity between learning sessions and find that the activity between successful and failed trials becomes more distinct. We also find evidence to support the idea of the formation of coherently active cell assemblies. Our results motivate further exploration of the differences in neural activity between failed and successful trials, particularly across learning sessions.

\subsection{Experimental Method}
\subsubsection{Experiment}
The experimental data used in this report comes from two sessions, spaced ten days apart, during which an initially naive monkey (\emph{rhesus macaque}) learned to perform a simple tongue protrusion task. During each trial the monkey's head was held in place with a tongue protrusive force transducer in its mouth, which it could press with its tongue to move a cursor on a screen up or down, the amount of force applied determining its position. The trial started with a $500ms$ hold period during which the monkey had to apply zero force to the transducer. Then a target rectangle above the cursor would appear. Once the monkey moved the cursor into the target rectangle it would have to hold it there for $200ms$. If the monkey succeeded in holding the cursor within the target during the hold period it would receive a small juice reward from the transducer. Each session lasted approximately one hour.

\subsubsection{Data Acquisition}
Neural activity was recorded by two microarrys of electrodes, one in the orofacial somatosensory (S1) cortex and the other in the motor (M1) cortex, cortices which are implicated in the control of tongue and jaw movements. These microarrays were not removed between sessions so most of the neurons recorded by the electrodes remained the same between the two days, though they have not been directly correlated. The force data from the transducer was also recorded. Neural data from these electrodes was preprocessed for spike candidates and then manually spike sorted using Offline Sorter from Plexon Inc. 

Neural activity was correlated with the timing of force onset. Unfortunately the algorithm for identifying the timing of force onset was imperfect and some trials had to be discarded and these were not the same trials in both cortices. Table \ref{table:n-trials-by-type} shows the number of trials of each type that were available available incorrectly-matched trials were removed. 

\begin{table}[h]
\begin{tabular}{|c|c|c|c|c|}
\hline 
 & \multicolumn{2}{c|}{M1} & \multicolumn{2}{c|}{S1} \\ 
\cline{2-5}
 & 11/07 & 11/17 & 11/07 & 11/17 \\ 
\hline 
Successes & 96 & 261 & 148 & 288 \\ 
\hline 
Overshoots & 21 & 114 & 25 & 118 \\ 
\hline 
Undershoots & 32 & 9 & 26 & 11 \\ 
\hline 
\end{tabular}
\caption{Numbers of events in each data set. While the M1 and S1 sets were recorded simultaneously we were unable to use the same trials due to difficulty in finding the time of movement onset. Thus, any comparison between the two is intended to be qualitative only.}\label{table:n-trials-by-type}
\end{table}

\section{Data Analysis}

\subsection{Neural Data}
\begin{figure}[h]
        \noindent\makebox[1\textwidth]{%
        \begin{subfigure}[b]{0.5\textwidth}
                \centering
                \includegraphics[width=\textwidth]{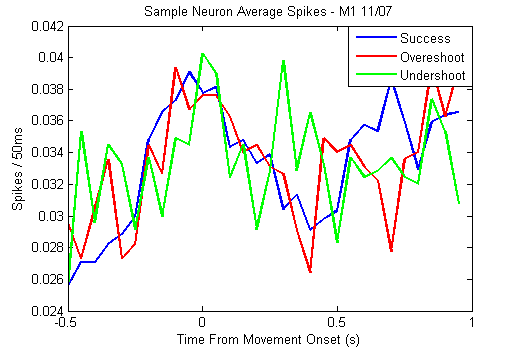}
                \caption{M1 - 11/07}
                \label{fig:samp-m1-07}
        \end{subfigure}
        \begin{subfigure}[b]{0.5\textwidth}
                \centering
                \includegraphics[width=\textwidth]{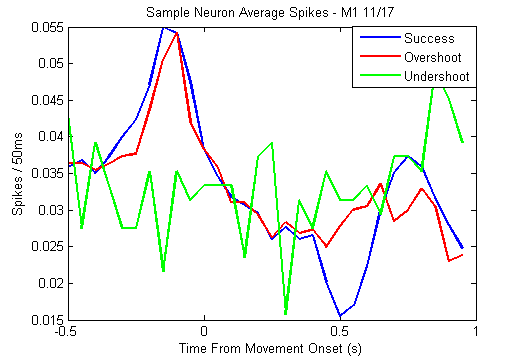}
                \caption{M1 - 11/17}
                \label{fig:samp-m1-17}
        \end{subfigure}}
        \\ \\
        \noindent\makebox[01\textwidth]{%
        \begin{subfigure}[b]{0.5\textwidth}
                \centering
                \includegraphics[width=\textwidth]{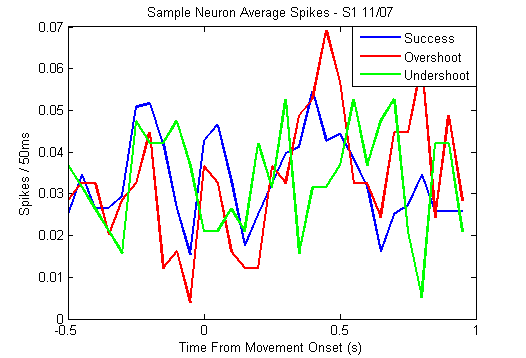}
                \caption{S1 - 11/07}
                \label{fig:samp-s1-07}
        \end{subfigure}
        \begin{subfigure}[b]{0.5\textwidth}
                \centering
                \includegraphics[width=\textwidth]{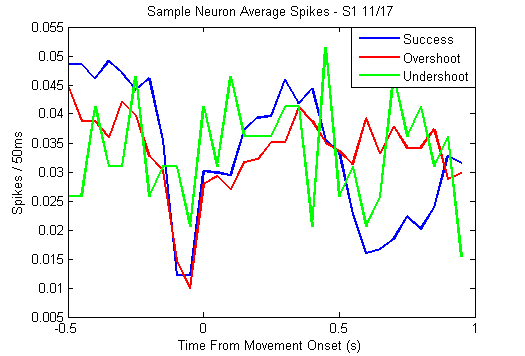}
                \caption{S1 - 11/17}
                \label{fig:samp-s1-17}
        \end{subfigure}}
        \\ \\
        \noindent\makebox[01\textwidth]{%
        \begin{subfigure}[b]{0.5\textwidth}
                \centering
                \includegraphics[width=\textwidth]{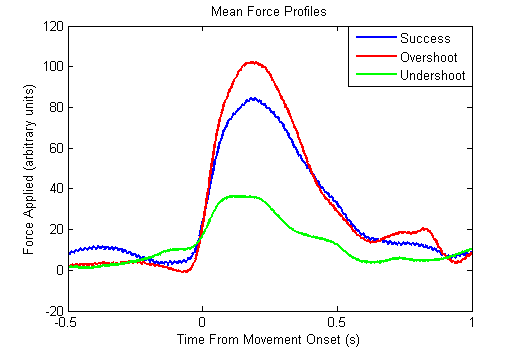}
                \caption{Force Profile - 11/07}
                \label{fig:force-s1-07}
        \end{subfigure}
        \begin{subfigure}[b]{0.5\textwidth}
                \centering
                \includegraphics[width=\textwidth]{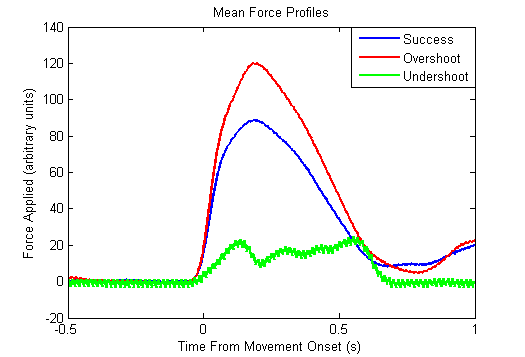}
                \caption{Force Profile - 11/17}
                \label{fig:force-s1-17}
        \end{subfigure}}
        \caption{Average firing rates for sample neurons in the S1 and M1 cortices during the two learning sessions and average force profiles for these the two sessions. Blue shows average activity during successful trials, red shows overshoots and green shows undershoots. The same neuron was used on each day.}\label{fig:samples}
\end{figure}

Figure \ref{fig:samples} displays average spike rate data from two sample neurons in the M1 and S1 cortices respectively on each day as well as average force profiles for those days. Each line shows the normalized average number of spikes in $50ms$ time bins over the time period from $-0.5s$ before force onset to $1.0s$ after for each of the three trial types. For these neurons, it is clear that they change their firing patterns very significantly between the two learning sessions and it appears that their patterns become more uniform for successful and overshoot trials. The undershot trials however, do not seem to show much of a pattern anywhere. This is partially due to the lower amount of data for undershoot trials and partially due to the nature of an undershoot trial, where the monkey may not move at all.

These single neurons, however, cannot tell us much about the ensemble activity or be taken as representative examples. Some neurons show very different patterns from these and some show no discernable firing pattern at all. However, it is useful to see individual examples before studying the behavior of the neurons on a whole.

\subsection{K-Nearest Neighbors Classification}

The first quantitative step in exploring the differences in neural activity between successful and failed trials is to find if there is enough difference to distinguish between them. To do this we used the K-Nearest Neighbors algorithm, a simple machine learning algorithm for the classification of data. For a given vector of test data it finds the $k$ closest training data vectors according to some distance metric. The test vector is then classified according to the category to which most of its nearest neighbors belong. We used the KNN algorithm with a $k$-value of 3 and a Euclidean distance metric to classify trials as either successful or a given failure type (success vs. overshoot and success vs. undershoot). As input data we used vectors of the activity (total spike count) of each neuron during a given $50~ms$ time window. This allowed us to compare classification performance as a function of time to see when differences in activity between trial types became pronounced. 

\begin{figure}[h]
        \noindent\makebox[1\textwidth]{%
        \begin{subfigure}[b]{0.75\textwidth}
                \centering
                \includegraphics[width=\textwidth]{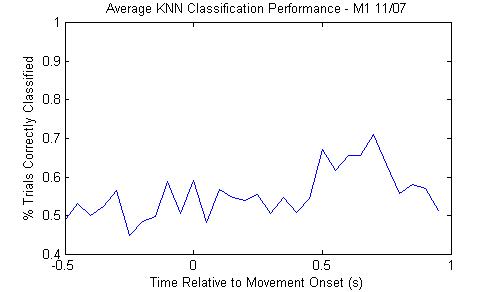}
                \caption{\large M1 - 11/07 \\}
                \label{fig:knn-m1-07}
        \end{subfigure}
        \begin{subfigure}[b]{0.75\textwidth}
                \centering
                \includegraphics[width=\textwidth]{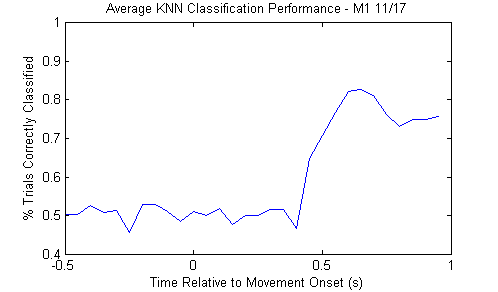}
                \caption{\large M1 - 11/17 \\}
                \label{fig:knn-m1-17}
        \end{subfigure}}
        \\ \\
        \noindent\makebox[01\textwidth]{%
        \begin{subfigure}[b]{0.75\textwidth}
                \centering
                \includegraphics[width=\textwidth]{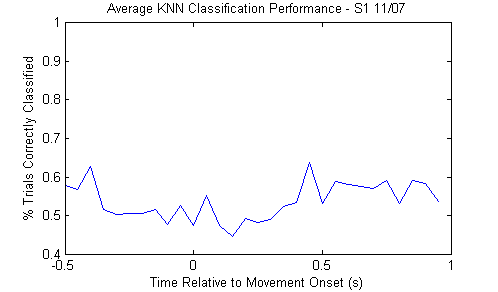}
                \caption{\large S1 - 11/07 \\}
                \label{fig:knn-s1-07}
        \end{subfigure}
        \begin{subfigure}[b]{0.75\textwidth}
                \centering
                \includegraphics[width=\textwidth]{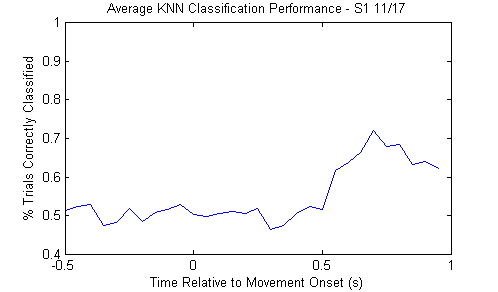}
                \caption{\large S1 - 11/17 \\}
                \label{fig:knn-s1-17}
        \end{subfigure}}
        \caption{Mean K-Nearest Neighbor classification performance for successful vs. overshoot trials. Classification is performed using spike rate data from each neuron during the given time bin. In both cortices, classification performs better on the second day and in all cases performs best half a second after movement onset. In both cortices, classification performance is also more uniform on the second day.}\label{fig:knn-21}
\end{figure}

We performed the classification a large number of times, randomly selecting trials for the training and sample data sets. This was necessary because the KNN algorithm requires an equal amount of training data from each category and it requires a large ratio of training to sample data. We don't have a lot of data for failed trials in general so this allowed us to obtain a better measure of performance. 

Figure \ref{fig:knn-21} shows the average classification performance (percentage of trials which are correctly classified) of the KNN algorithm when distinguishing between successful and overshoot trials. The left column shows the results from the first day and the right column shows the results from the second day. In both cortices the classifier performed no better than chance until half a second after movement onset. It is also clear that in both cortices the classifier's performance on the second day's data is significantly improved from the previous day. This is probably due in part to the larger amount of data available during the second day for overshoot trials, but this result motivates further examination of the neural data to see how it changes from day to day for successes and failures.

\begin{figure}[h]
        \noindent\makebox[1\textwidth]{%
        \begin{subfigure}[b]{0.75\textwidth}
                \centering
                \includegraphics[width=\textwidth]{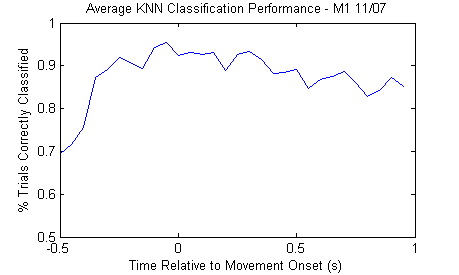}
                \caption{\large M1 - 11/07 \\}
                \label{fig:knn-m1-07-24}
        \end{subfigure}
        \begin{subfigure}[b]{0.75\textwidth}
                \centering
                \includegraphics[width=\textwidth]{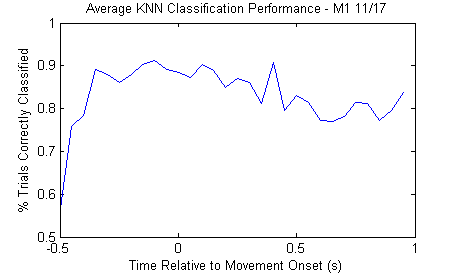}
                \caption{\large M1 - 11/17 \\}
                \label{fig:knn-m1-17-24}
        \end{subfigure}}
        \\ \\
        \noindent\makebox[01\textwidth]{%
        \begin{subfigure}[b]{0.75\textwidth}
                \centering
                \includegraphics[width=\textwidth]{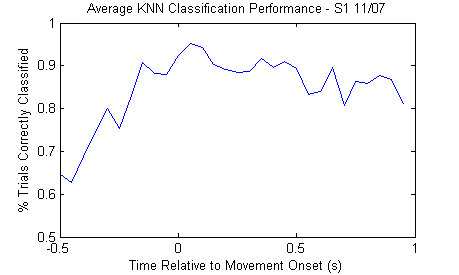}
                \caption{\large S1 - 11/07 \\}
                \label{fig:knn-s1-07-24}
        \end{subfigure}
        \begin{subfigure}[b]{0.75\textwidth}
                \centering
                \includegraphics[width=\textwidth]{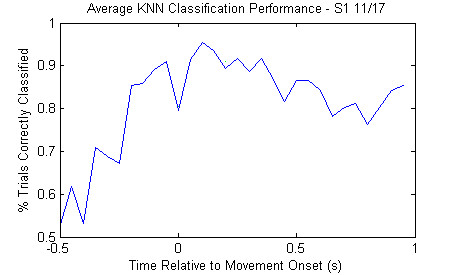}
                \caption{\large S1 - 11/17 \\}
                \label{fig:knn-s1-17-24}
        \end{subfigure}}
        \caption{Mean K-Nearest Neighbor classification performance for successful vs. undershoot trials. Classification is performed using spike rate data from each neuron during the given time bin. Classification performance improves before movement onset and remains high in all cases, but the jump is slightly delayed in the S1 cortex as compared to the M1 cortex.}\label{fig:knn-24}
\end{figure}

Figure \ref{fig:knn-24} shows the performance of the classifier when distinguishing between successful and undershoot trials. In all cases the classifier performance is better than chance at all times and jumps to above $90\%$ shortly before movement onset. This jump is slightly delayed in the S1 cortex, particularly on the second day, which may be an effect of the small data sample. However, it is clear from what we've seen of activity during undershoot trials that no consistent pattern appears in the neural activity. It is this lack of a pattern which makes this failure type so easily distinguishable from a success or failure. It seems there is nothing particularly interesting to learn about activity during undershoot trials so they will be ignored for the remainder of this report, but future studies might look further into the force profiles of these trials to find a better method of analysis.

These results demonstrate that neural activity in the M1 and S1 cortices is associated with the outcome of the trial. In the case of overshoot trials it seems that the main difference in activity occurs in the feedback period, several hundred milliseconds after the reward has been dispensed. This difference in activity may be solely due to the motion of swallowing the juice following successful trials, but the improvement in classification performance on the second day of trials suggests that neurons may be responding more consistently, indicating a difference due to learning.

\subsection{Signrank}

The previous section was not able to demonstrate the hypothesis that neural activity patterns changed between days of testing because the larger amount of data available for the KNN algorithm may have been sufficient to cause the observed increase in classification performance. We test the hypothesis by more directly examining the differences in activity of neurons between successful and overshoot trials on each day of testing. If this hypothesis is true we expect more neurons to differentiate their activity based on the success or failure of a trial on the second day than on the first.

To numerically evaluate the difference in activity we compare the mean firing rates of each neuron during successful and overshoot trials using a two-sided Wilcoxon signed rank test. This algorithm determines the probability $p$ that the difference between the two input vectors describes a distribution with a mean value of zero. We first find the average number of spikes produced by each neuron in $25ms$ time bins for each trial type. We then divide the trials into two epochs, the feedforward epoch ($-0.5s \leq t < 0.25s$), during which the neurons have no feedback from the trial success, and the feedback epoch ($0.25 \leq t < 1.0s$), which takes place place after the trial has failed or succeeded. For each epoch we perform a signed rank test on the first and last $500ms$ of the epoch (overlapping by $250ms$). If either test determines that the difference in average activity is significant ($p<0.01$) the neuron is classified as either a feedback or feedforward-modulated neuron. 

\begin{table}[h]
\begin{tabular}{|c|c|c|c|c|}
\hline 
 & \multicolumn{2}{c|}{M1} & \multicolumn{2}{c|}{S1} \\ 
\cline{2-5}
 & 11/07 & 11/17 & 11/07 & 11/17 \\ 
\hline 
Total Neurons & 121 & 118 & 110 & 118 \\ 
\hline 
Feedforward-Modulated Neurons & 14 & 7 & 18 & 5 \\ 
\hline 
Feedback-Modulated Neurons & 34 & 75 & 25 & 35 \\ 
\hline 
\end{tabular} 
\caption{Numbers of neurons which showed a significant difference ($p<0.01$) in average firing rates for overshoot trials compared to successful trials, as calculated with a two-sided Wilcoxon signed rank test. The first epoch studied was the time period from -0.5 to 0.25 seconds relative to movement onset and the second epoch was the period from 0.25 to 1.0 seconds relative to movement onset.}\label{table:signrank}
\end{table}

Table \ref{table:signrank} shows the numbers of neurons which showed a significant difference ($p<0.01$) in average firing rates for overshoot trials as compared to successful ones. In both cortices the number in the first epoch decreased from the first day of trials to the second and the number in the second epoch increased. This suggests that preparatory activity became more uniform for both successful and failed trials while the activity after motion differentiated into different specific patterns depending on the success or failure of the trial. The change is most marked for feedback-modulated neurons in the M1 cortex, the number of which more than doubles from the first day to the second. This is in line with the findings of the KNN classifier, which showed a performance improvement between days for this epoch and suggests that this improvement was not due to the larger sample size available to the classifier.

\begin{figure}[h]
\includegraphics[width=\textwidth]{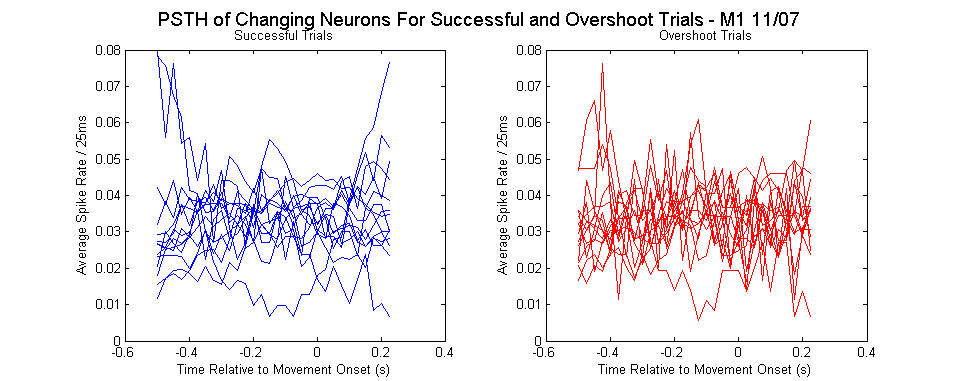}
\caption{Peri-Stimulus Time Histogram for neurons which show different activity for successful versus failed trials. Each plot shows the average activity of each neuron across the respective trial types.}\label{plot:psth-07-pre}
\end{figure}
\begin{figure}
\includegraphics[width=\textwidth]{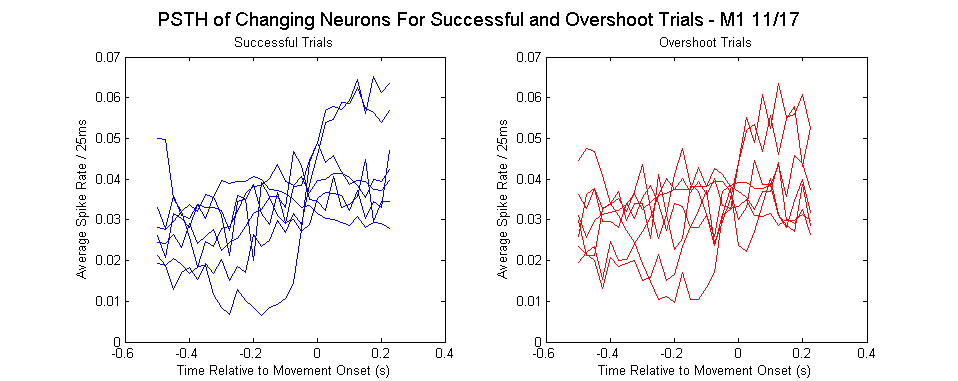}
\caption{Peri-Stimulus Time Histogram for neurons which show different activity for successful versus failed trials. Each plot shows the average activity of each neuron across the respective trial types. Variability in activity has decreased for this epoch since the first day of trials.}\label{plot:psth-17-pre}
\end{figure}
\begin{figure}[h]
\includegraphics[width=\textwidth]{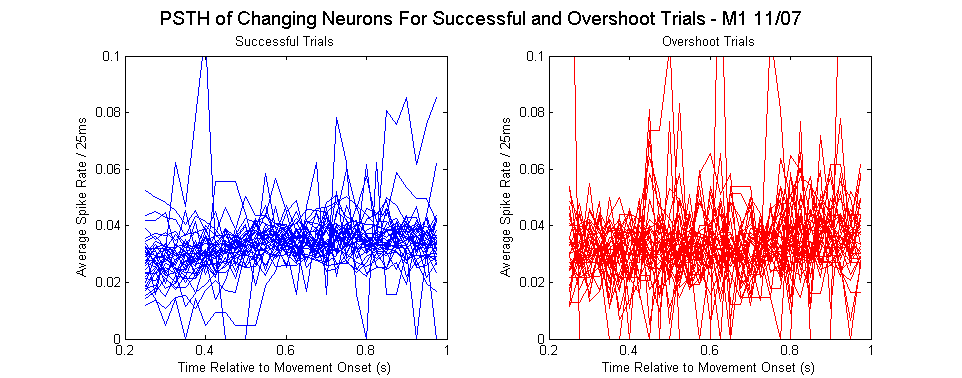}
\caption{Post-Stimulus Time Histogram for neurons which show different activity for successful versus failed trials on the first day. Each plot shows the average activity of each neuron across the respective trial types.}\label{plot:psth-07-pos}
\end{figure}
\begin{figure}[h]
\includegraphics[width=\textwidth]{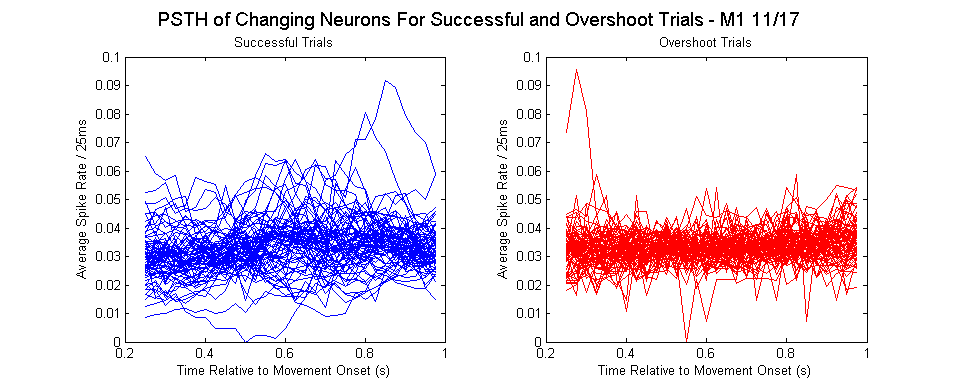}
\caption{Post-Stimulus Time Histogram for neurons which show different activity for successful versus failed trials on the second day. Each plot shows the average activity of each neuron across the respective trial types. Activity has become more differentiated between trial types since the first day. Successful trials are associated with specific spiking patterns while overshoot trials are associated with little to no modulation in activity.}\label{plot:psth-17-pos}
\end{figure}

Figures \ref{plot:psth-07-pre} through \ref{plot:psth-17-pos} show the mean firing rates (normalized for comparison) of feedforward-modulated neurons during successful and overshoot trials, with successful trials in blue and failed trials in red. The change in neural activity from the first day to the second is most marked in the feedback epoch, where clear patterns emerge following successful trials and activity becomes much more uniform following unsuccessful trials. 

\begin{figure}[h]
        \noindent\makebox[1\textwidth]{%
        \begin{subfigure}[b]{0.75\textwidth}
                \centering
                \includegraphics[width=\textwidth]{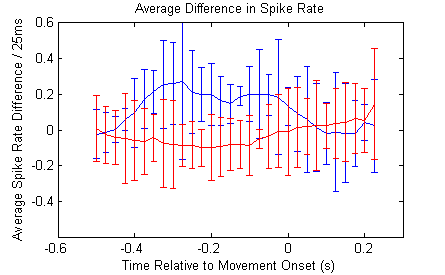}
                \caption{\large Feedforward - 11/07 \\}
                \label{fig:diffs-pre-m1-07}
        \end{subfigure}
        \begin{subfigure}[b]{0.75\textwidth}
                \centering
                \includegraphics[width=\textwidth]{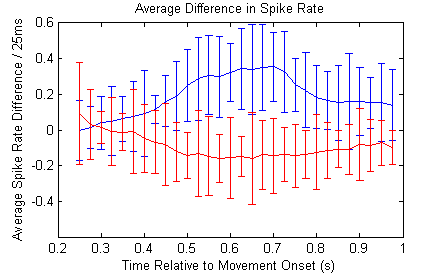}
                \caption{\large Feedback - 11/07 \\}
                \label{fig:diffs-pos-m1-07}
        \end{subfigure}}
        \\ \\
        \noindent\makebox[01\textwidth]{%
        \begin{subfigure}[b]{0.75\textwidth}
                \centering
                \includegraphics[width=\textwidth]{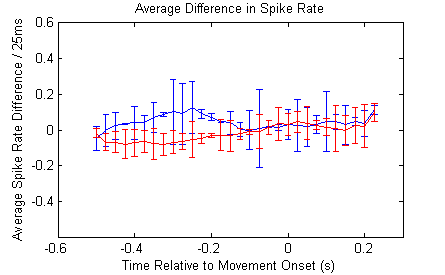}
                \caption{\large Feedforward - 11/17 \\}
                \label{fig:diffs-pre-m1-17}
        \end{subfigure}
        \begin{subfigure}[b]{0.75\textwidth}
                \centering
                \includegraphics[width=\textwidth]{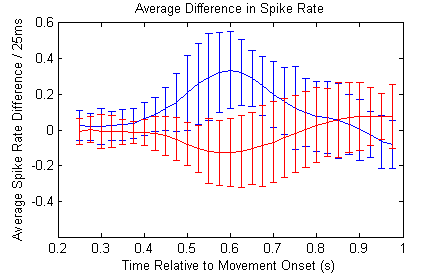}
                \caption{\large Feedback - 11/17 \\}
                \label{fig:diffs-pos-m1-17}
        \end{subfigure}}
        \caption{M1 Cortex data for high and low-PCA score neuron group averages of the difference in activity for successful and overshoot trials. Activity becomes more uniform on the second day, particularly in the feedforward epoch. In the feedback epoch the timing of the difference in activity occurs at a more consistent time across neurons.}\label{fig:diffs-m1}
\end{figure}

\begin{figure}[h]
        \noindent\makebox[1\textwidth]{%
        \begin{subfigure}[b]{0.75\textwidth}
                \centering
                \includegraphics[width=\textwidth]{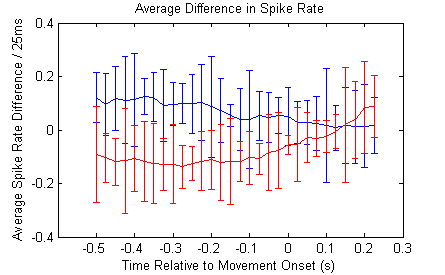}
                \caption{\large Feedforward - 11/07 \\}
                \label{fig:diffs-pre-s1-07}
        \end{subfigure}
        \begin{subfigure}[b]{0.75\textwidth}
                \centering
                \includegraphics[width=\textwidth]{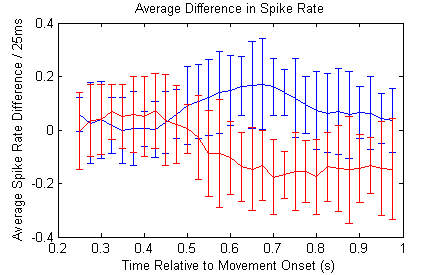}
                \caption{\large Feedback - 11/07 \\}
                \label{fig:diffs-pos-s1-07}
        \end{subfigure}}
        \\ \\
        \noindent\makebox[01\textwidth]{%
        \begin{subfigure}[b]{0.75\textwidth}
                \centering
                \includegraphics[width=\textwidth]{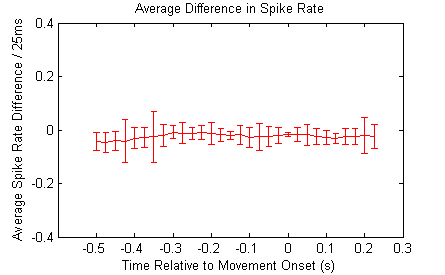}
                \caption{\large Feedforward - 11/17 \\}
                \label{fig:diffs-pre-s1-17}
        \end{subfigure}
        \begin{subfigure}[b]{0.75\textwidth}
                \centering
                \includegraphics[width=\textwidth]{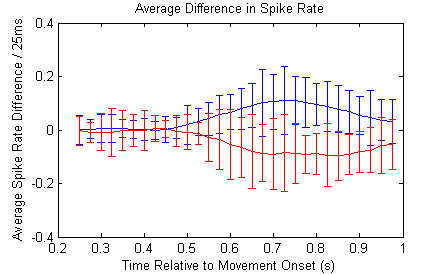}
                \caption{\large Feedback - 11/17 \\}
                \label{fig:diffs-pos-s1-17}
        \end{subfigure}}
        \caption{S1 Cortex data for high and low-PCA score neuron group averages of the difference in activity for successful and overshoot trials. Activity becomes more uniform on the second day, particularly in the feedforward epoch.}\label{fig:diffs-s1}
\end{figure}

To illustrate this more clearly we took the differences in the activity of each neuron and used Principal Component Analysis (PCA) to loosely categorize the neurons based on when they differ and by how much. We then grouped them by the first component of their PCA scores, selecting those with high or low scores (beyond 0.25 standard deviations from the mean PCA score). Figures \ref{fig:diffs-m1} and \ref{fig:diffs-s1} show plots of the averages of these two groups with errorbars corresponding to one standard deviation.

The change in the feedforward activity in both cortices between the learning sessions is very noticeable. There is a very clear trend towards homogeneity; fewer neurons show a difference between successful and overshoot trials and the differences are far less substantial. The feedforward epochs show a subtler change. The magnitude of the average differences of these outlying groups does not change significantly, but they become more consistent in time. Looking back at figures \ref{plot:psth-07-pos} and \ref{plot:psth-17-pos}, the reasons for this seem to be that on the first day, activity in both successful and overshoot trials changed a lot and was inconsistent, while on the second day, successful trials began to show a pattern of modulation and the overshoot trials modulated a lot less. Thus, the magnitude of difference remained the same despite markedly different neural activity. Part of the smoothing out that happens is because the second day had more data on these trials, but in both cortices it's clear that the activity becomes more coherent.

\section{Discussion and Conclusions}
This report was a first look at the differences in neural activity between successful and failed trials. We found that a significant difference in neuron firing patterns for successful and overshoot trials appeared during the second learning session; neurons began to fire in more consistent patterns, modulating after successful trials and not modulating after over shoot trials, and seemed to form groups of coherently active cells, consistent with the findings of Arce \emph{et al}\cite{farce-adaptation}. Preparatory activity became more uniform and less differentiated between successful and overshoot trials. Undershoot trials did not show any consistent pattern, but deeper analysis may be able to find some regularities, perhaps by improving the force onset timing for these trials.

Our results motivate further exploration into the differences in activity between successful and unsuccessful trials, particularly with overshoot trials. Obtaining an exact correlation of cells across learning sessions would enable a more in-depth look at the formation of coherently active assemblies as the monkey learns. Analysis could also focus on shifts in activity within learning sessions to identify changes that take place between sessions as compared to during sessions. Finally, slight differences in the timing of activity in the M1 and S1 cortices observed in previous experiments and this study suggests that activity in one cortex might influence the other. Future studies might attempt to find correlations in the activity of coherently active cell clusters between the two cortices to look for any sort of feedback effects.

\subsection{Acknowledgments}
The authors would like to thank Dr. Nicholas Hatsopoulos, Dr. Deborah Nelson and the Hatsopoulos lab for support and consultation. This undergraduate research project was supported by Dr. Nicholas Hatsopoulos and supervised by Dr. Fritzie Arce.

\bibliographystyle{plain}
\bibliography{biblio}

\begin{thebibliography}{1}

\bibitem{farce-adaptation}
F.~Arce, C.F. Rossi, J.C. Lee, B.J. Sessle, and N.G. Hatsopoulos.
\newblock Dynamics of modulation in orofacial sensorimotor cortex during
  long-term adaptation to a novel tongue-protrusion task.
\newblock In N.G. Hatsopoulos, editor, {\em •}, 2011.

\end{thebibliography}

\end{document}